\documentclass[aps,prx,superscriptaddress,reprint, longbibliography]{revtex4-2}
\usepackage[usenames,dvipsnames]{xcolor}
\usepackage{amsmath}
\usepackage{amssymb}
\usepackage{graphicx}
\usepackage{url}
\usepackage{hyperref}
\usepackage{color,xcolor}
\usepackage{epstopdf}
\usepackage{float}
\usepackage{ulem}
\usepackage[percent]{overpic}
\usepackage{siunitx}
\usepackage{subdepth}
\usepackage{enumitem}

\usepackage{dcolumn}
\usepackage{bm}
\usepackage{times}
\usepackage[version=3]{mhchem} 

\usepackage{mathrsfs}
\usepackage{lipsum}
\usepackage{amsfonts}
\usepackage{array}
\usepackage{ulem}
\usepackage{lineno}
\usepackage[none]{hyphenat}

\hypersetup{colorlinks=true,citecolor=Red,linkcolor=Blue,urlcolor=Black}

\makeatletter
\def\maketitle{
	\@author@finish
	\title@column\titleblock@produce
	\suppressfloats[t]}
\makeatother

\begin{document}

\title{Kinetic obstruction to pairing in the doped Kitaev-Heisenberg ladder}

\author{Bradraj Pandey}\email{bradraj.pandey@gmail.com}
\affiliation{Department of Physics and Astronomy, University of Missouri, Columbia, Missouri 65211, USA}
\affiliation{Department of Physics and Astronomy, University of Tennessee, Knoxville, Tennessee 37996, USA}

\author{Bo Xiao}
\affiliation{Materials Science and Technology Division, Oak Ridge National Laboratory, Oak Ridge, Tennessee 37831, USA}
\affiliation{Quantum Science Center, Oak Ridge, Tennessee 37831, USA}

\author{Satoshi Okamoto}
\affiliation{Materials Science and Technology Division, Oak Ridge National Laboratory, Oak Ridge, Tennessee 37831, USA}

\author{Gonzalo Alvarez}
\affiliation{Computational Sciences and Engineering Division, Oak Ridge National Laboratory, Oak Ridge, Tennessee 37831, USA}

\author{G\'abor B. Hal\'asz}
\affiliation{Materials Science and Technology Division, Oak Ridge National Laboratory, Oak Ridge, Tennessee 37831, USA}
\affiliation{Quantum Science Center, Oak Ridge, Tennessee 37831, USA}

\author{Elbio Dagotto}
\affiliation{Department of Physics and Astronomy, University of Tennessee, Knoxville, Tennessee 37996, USA}
\affiliation{Materials Science and Technology Division, Oak Ridge National Laboratory, Oak Ridge, Tennessee 37831, USA}

\author{Pontus Laurell}\email{plaurell@missouri.edu}
\affiliation{Department of Physics and Astronomy, University of Missouri, Columbia, Missouri 65211, USA}
\affiliation{Materials Science and Engineering Institute, University of Missouri, Columbia, Missouri 65211, USA}

\begin{abstract}
We investigate the hole-doped Kitaev-Heisenberg ($t$--$J$--$K$) model on a two-leg ladder geometry using the density-matrix renormalization group (DMRG). We first consider the behavior of the antiferromagnetic Kitaev (AFK) spin-liquid phase as a function of hopping strength $t$ and doping level. This reveals intriguing pairing tendencies only for $\frac{t}{K} \lesssim 0.65$, consistent with prior results on three-leg ladders, and firmly supports the emerging picture that the physics of doped Kitaev spin liquids strongly depends on the kinetic energy of the doped holes. Analysis of one- and two-hole doping uncovers close links between the spatial profiles of the plaquette operator and the charge density.
We construct a doping-dependent phase diagram for antiferromagnetic Heisenberg interactions and intermediate hopping $t=1$. Upon doping, the rung-singlet region develops dominant superconducting correlations. 
Charge-density-wave correlations dominate at weak doping near the transition to the stripy phase.
Spin-density wave-like behavior is found in the AFK and ferromagnetic Kitaev limits, and in the stripy phase.
\end{abstract}
\date{March 12, 2026}

\maketitle

\noindent {\bf Introduction} \\
The response of quantum spin liquids (QSLs) to charge doping presents many open questions. QSLs are intriguing exotic phases of matter found in strongly interacting spin systems, characterized by long-range entanglement, fractionalized excitations, and the absence of magnetic ordering even at zero temperature \cite{Savary2017, Knolle2019, Broholm2020}. However, their strong interactions and high entanglement present substantial challenges for theoretical and computational modeling. The exactly solvable Kitaev spin liquid \cite{Kitaev2006} has thus attracted particular interest as it allows for rigorous results. This model consists of bond-dependent Ising interactions on a honeycomb lattice, and hosts a QSL ground state featuring gapless Majorana and gapped $Z_2$ flux (vison) excitations \cite{Kitaev2006, Hermanns2018, Mandal_2025_Kitaev}. In the presence of a magnetic field, the system undergoes a transition to a topological chiral QSL phase that supports Ising non-Abelian anyons, providing a potential platform for fault-tolerant topological quantum computation \cite{Kitaev2006, RevModPhys.80.1083}.

Motivated by Kitaev's honeycomb model, a number of spin-orbit-coupled transition metal compounds \cite{PhysRevLett.102.017205, Takagi2019, Trebst2022, Mandal_2025_Kitaev, Matsuda2025} have been explored as candidate materials, notably including Na$_2$IrO$_3$ \cite{PhysRevLett.108.127203, PhysRevLett.108.127204, Chun2015} and $\alpha$-RuCl$_3$ \cite{PhysRevB.90.041112, Banerjee2017, Banerjee2018, Laurell2020, Wang2020a, Moeller2025RuCl3}. However,  
additional non-Kitaev interactions, such as Heisenberg exchange terms, destabilize the Kitaev QSL phase, resulting in a magnetically ordered ground state in most candidate materials, and make the model non-integrable. While certain aspects can nevertheless be explored using perturbative or mean-field methods, numerical methods are often required for an accurate description of the properties away from equilibrium or half-filling. Tensor network approaches~\cite{Schollwock2011, Orus2019, Xiang2023} provide the state-of-the-art framework for this regime, most notably by utilizing the density matrix renormalization group (DMRG)~\cite{PhysRevLett.69.2863, PhysRevB.48.10345} to variationally optimize the matrix product state ansatz~\cite{PhysRevLett.119.157203, Jiang2019, Xiao2025}.

It has been proposed that doping the Kitaev QSL can give rise to exotic superconducting states \cite{PhysRevB.85.140510, PhysRevLett.108.227207, PhysRevB.86.085145, PhysRevB.87.064508, PhysRevB.90.045135, PhysRevB.97.014504} and could produce insights into the mechanisms of high-temperature superconductivity in cuprates \cite{RevModPhys.66.763, RevModPhys.78.17}. However, recent DMRG studies on three- and four-leg cylinders have shown that, in the pure Kitaev limit, the superconducting pair–pair correlation functions decay exponentially for large hopping amplitudes (corresponding to the so-called fast-hole regime) \cite{Peng2021, Jin2024, PhysRevB.110.224518}. In fact, our recent study \cite{PhysRevB.110.224518} shows that the binding of two holes occurs only at low hopping strengths, i.e. in the slow-hole regime, supporting a distinction between the physics of slow and fast holes in the Kitaev QSL. It has also recently been demonstrated that, upon single-hole doping, signatures of Nagaoka-type kinetic ferromagnetism emerge in the ferromagnetic Kitaev (FK) QSL phase \cite{Jin2024}, whereas in the antiferromagnetic (AFK) QSL phase, the single-hole spectral function exhibits characteristics of fractionalized excitations \cite{Kadow_2024}, reflecting the intricate interplay between charge, spin, and flux degrees of freedom. Similar findings were also reported on three-leg cylinders for the triangular lattice \cite{Hardy2025kitaevSC}.

This recent progress notwithstanding, questions remain as to what happens when the FK and AFK models are doped, both microscopically and on a larger scale, i.e. which phases do emerge? Prior studies have focused on cylindrical multi-leg geometries, which are associated with large computational costs and challenges for finite-size scaling. In this work, we instead investigate the doped Kitaev--Heisenberg model on a two-leg ladder, with a particular focus on pairing tendencies in the Kitaev spin-liquid, rung-singlet, and stripy phases. This quasi-one-dimensional geometry is numerically more tractable, allowing us to obtain more robust results and to map out a phase diagram. In addition, we note that the two-leg ladder geometry has previously been found to yield results in surprisingly close correspondence to the two-dimensional (2D) limit \cite{PhysRevB.99.195112, PhysRevB.99.224418, PhysRevX.11.011013}. Thus, much like with models for cuprate superconductors \cite{Dagotto_1999}, we expect that reliable ladder results are informative about the physics also in the 2D limit.

The phase diagram of the Kitaev-Heisenberg ladder at half-filling is known \cite{PhysRevB.99.195112, PhysRevB.99.224418, PhysRevB.108.045124, Sousa2025KH}, but the doped model has yet to be explored. Here we obtain the ground-state properties and analyze correlation functions together with the plaquette operator using DMRG methods. 
Our results show that pairing tendencies in the spin-liquid phases are strongly controlled by the kinetic energy of the doped holes. In the AFK region, pairing between two holes persists only for hopping strengths $\frac{t}{K} \lesssim 0.65$, beyond which spin-density-wave order develops. In contrast, the FK region is much more fragile: pairing survives only for very small hopping values $\frac{t}{K} \lesssim 0.1$, and further increase in $t$ subsequently leads to Nagaoka-type ferromagnetic order.

We also reveal signatures of this kinetic obstruction to pairing in the plaquette operator: (i) its spatial profile mimics the charge distribution and thus depends on whether holes form pairs or repel each other, and (ii) the slope of its spatial average changes at the hopping strengths where pairing disappears. 
Furthermore, we construct a doping-dependent phase diagram at intermediate hopping $t=1$ by varying the Kitaev and Heisenberg interactions. We find that superconducting correlations appear only in the rung-singlet phase, whereas charge-density-wave order emerges at weak doping near the transition between the rung-singlet and stripy phases. Other regions in the phase diagram develop spin-density wave-like behavior. Overall, our results uncover a rich interplay among pairing, charge, flux, and spin orders in the doped Kitaev–Heisenberg ladder.

\begin{figure*}[!ht]
\hspace*{-0.5cm}
\vspace*{0cm}
\begin{overpic}[width=2.1\columnwidth]{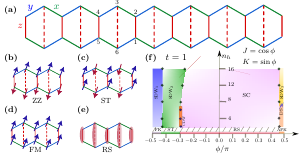}
\end{overpic}
\caption{{\bf Kitaev–Heisenberg model and doping-dependent phase diagram.}  
{\bf (a)} Schematic of the two-leg honeycomb ladder. The three inequivalent bonds, $x$, $y$, and $z$, are shown as solid green, blue, and red lines, respectively. Dashed red lines indicate $z$-type bonds connecting sites across the periodic boundary of the ladder.  
{\bf (b)}–{\bf (e)} Pictorial representations of the {\bf (b)} zigzag (ZZ), {\bf (c)} stripy (ST), {\bf (d)} ferromagnetic (FM), and {\bf (e)} rung-singlet (RS) magnetic phases occurring at half-filling. Ellipses in {\bf (e)} indicate singlet states formed by the two enclosed spins. 
{\bf (f)} Doping-dependent phase diagram obtained by parametrizing the interactions as $J=\cos\phi$ and $K=\sin\phi$ over the interval $-\pi/2 \le \phi \le \pi/2$, with fixed intermediate hopping $t=1$. The phase diagram at half-filling is shown at the bottom, below the hatched area. Upon finite doping (above the hatched area), the rung-singlet phase evolves into a superconducting (SC) phase, whereas the ferromagnetic Kitaev (FK) phase evolves into a commensurate spin-density-wave (SDW) order SDW$_1$ and the stripy phase evolves into an incommensurate SDW$_2$. A narrow range dominated by charge-density wave (CDW) correlations is also found near the ST-RS transition. At low doping levels in the AFK limit a potential disorder phase (DS) is identified. At higher doping levels it evolves into an incommensurate SDW$_3$ phase.}
\label{fig1}
\end{figure*}

\noindent {\bf \\Results\\}
\noindent {\small \bf Model Hamiltonian\\}
The $t$--$J$--$K$ Hamiltonian for the doped Kitaev--Heisenberg model \cite{PhysRevB.85.140510, PhysRevB.86.085145, PhysRevB.87.064508, PhysRevLett.111.037205, PhysRevB.90.024404, PhysRevB.90.045135, PhysRevB.110.224518} is given by
\begin{align}
    H &= K \sum_{\langle i,j \rangle} S_i^{\gamma} S_j^{\gamma}
       + J \sum_{\langle i,j \rangle} \mathbf{S}_i \cdot \mathbf{S}_j \nonumber\\
      &\quad - t \sum_{\langle i,j \rangle,\sigma}
      \mathcal{P} \left( c^{\dagger}_{i\sigma} c_{j\sigma} + \mathrm{H.c.} \right) \mathcal{P},
    \label{eq:ham}
\end{align}
where $K = \sin \phi$ is the Kitaev interaction, $J = \cos \phi$ is the Heisenberg exchange, and $t$ is the hopping amplitude of the doped holes.
Here, $\gamma$ labels the type of bond connecting sites $i$ and $j$, $S_i^{\gamma}$ is the $\gamma \in \{x,y,z\}$ component of the spin-$1/2$ operator, and
$\mathbf{S}_i = (S_i^x, S_i^y, S_i^z)$ is the spin vector on site $i$.
The summation $\sum_{\langle i,j \rangle}$ runs over nearest neighbors (NN), and $c_{i\sigma}^\dagger$ is the fermionic creation operator. The projector $\mathcal{P}$ imposes the constraint of no double occupancy, typical of $t$--$J$--like models. 
We solve Eq.~\eqref{eq:ham}  on a two-leg ladder numerically using the DMRG method \cite{PhysRevLett.69.2863, PhysRevB.48.10345} leveraging the computational frameworks of DMRG++ \cite{Alvarez2009} and ITensor \cite{itensor, itensor-r0.3}. We employ a two-leg ladder geometry, as illustrated in Fig.~\ref{fig1}(a) with $N = 2L_x$ sites and open boundary conditions (OBC) along the leg direction, whereas periodic boundary conditions (PBC) are applied along the rung direction so that most sites have a coordination number of three \cite{PhysRevB.99.195112, PhysRevB.99.224418}. 
In the following we will report results for both fixed $t=1$ (when varying $\phi$) and varying $\frac{t}{|K|}$ (when keeping $\phi=\pm \frac{\pi}{2}$ fixed so that $|K|=1$).

\begin{figure*}[!ht]
\hspace*{-0.5cm}
\vspace*{0cm}
\begin{overpic}[width=2.1\columnwidth]{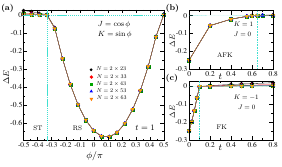}
\end{overpic}
\caption{{\bf Hole-pair formation.}  
{\bf (a)} Binding energy $\Delta E$ for different system sizes $N$, obtained using DMRG for $-\pi/2 \le \phi \le \pi/2$ with fixed hopping $t=1$, where the couplings are parametrized as $J=\cos\phi$ and $K=\sin\phi$. At this intermediate hopping strength, pair formation is found only in the rung singlet phase. 
{\bf (b)} Binding energy $\Delta E$ as a function of hopping $t$ in the AFK limit ($K=1$, $J=0$) for different system sizes $N$.  
{\bf (c)} Binding energy $\Delta E$ as a function of $t$ in the FK limit ($K=-1$, $J=0$) for different $N$.  
In both the AFK and FK regimes, the vertical dashed cyan line marks the hopping strength at which the binding energy becomes non-negative, occurring at $t^c_\mathrm{AFK} \approx 0.65K$ and $t^c_\mathrm{FK} \approx 0.1|K|$, respectively.}
\label{fig2}
\end{figure*}

\noindent {\small \bf Phase Diagram\\}
The ground-state phase diagram at half filling (i.e., one electron per site) as a function of $\phi$ exhibits six distinct phases \cite{PhysRevB.99.195112, PhysRevB.99.224418}. Among them, four are magnetically ordered: (i) a zigzag phase with antiferromagnetic (AFM) ordering along the rungs and ferromagnetic (FM) alignment along the legs [Fig.~\ref{fig1}(b)]; (ii) a stripy phase characterized by FM rung alignment and AFM leg ordering [Fig.~\ref{fig1}(c)]; (iii) a ferromagnetic phase with spin alignment predominantly in the $XY$ plane [Fig.~\ref{fig1}(d)]; and (iv) a rung-singlet phase featuring spin singlets formed along the rungs [Fig.~\ref{fig1}(e)]. The rung-singlet phase is the ladder counterpart to the N\'eel AFM phase in the 2D lattice. In addition to these ordered states, we identify the (v) AFK and (vi) FK spin-liquid phases emerging near $\phi = \pi/2$ and $-\pi/2$, respectively.  The phase boundaries are determined from the spin–spin correlation functions and singular features in the second derivative of the ground-state energy, $\chi_E = -\partial^2_\phi E_0$ (see Supplementary Note 1).

In Fig.~\ref{fig1}(f), we present the phase diagram as a function of hole doping ($n_h$) and interaction angle $\phi$ ($-0.5\pi \leq \phi \leq 0.5\pi$) for a fixed hopping amplitude $t = 1$ and system size $N = 126$. (Four of the six distinct phases at half-filling are found in this range of $\phi$, and thus displayed in Fig.~\ref{fig1}(f).) To identify the various phases, we compute the binding energy, spin–spin, charge–charge, and both singlet and triplet pair–pair correlation functions. Upon doping, the rung-singlet phase develops dominant singlet superconducting correlations, accompanied by competing charge-density correlations and exponentially decaying spin–spin correlations. In contrast, in the AFK limit at weak doping ($n_h<8$) we observe a potentially disordered phase (DS). In this regime, both spin-spin an charge-charge correlations decay with similar power-law behavior. However the static spin structure factor exhibits broad features, reminiscent of the behavior at half-filling, suggesting a QSL-like background lacking clear ordering wave vector signals. As the doping level is increased further, spin-spin correlations become dominant, and the static spin and charge structure factors develop incommensurate peaks; we denote this as the incommensurate spin-density-wave (SDW$_3$) phase (see Supplementary Note 2).

In the FK regime, however, even weak hole doping enhances the spin–spin correlations from an exponential decay to nearly long-range behavior at higher doping levels. This strong ferromagnetic response originates from the formation of a Nagaoka-type ferromagnet (SDW$_1$) induced by mobile holes. 
The stripy phase evolves into an incommensurate spin-density-wave magnetic order (SDW$_{2}$) that persists across all studied doping strengths. Near the SDW$_{2}$-SC boundary at lower hole concentrations ($2 \le n_h < 8$), we find intertwined charge, spin, and superconducting orders \cite{PhysRevB.103.214513}, where all corresponding correlations decay close to power-law behavior. In this region, we find that charge-charge correlations dominate, which we identify as the charge-density-wave (CDW) phase (see Supplementary Note 2). The pair-pair correlations decay exponentially in all phases except for the rung-singlet phase, indicating the absence of superconducting order outside the rung-singlet regime.

\noindent {\small \bf Pairing tendencies\\}
To investigate the pairing tendencies of holes in the Kitaev–Heisenberg ladder, we calculate the binding energy $\Delta E$, defined near half filling as \cite{PhysRevB.103.214513}
\begin{equation}
\Delta E = E(N-2) + E(N) - 2E(N-1),
\label{eq:binding}
\end{equation}
where $E(n_e)$ denotes the ground-state energy with $n_e$ electrons. The hole concentration is quantified as $n_h = N - n_e$, representing the number of holes introduced into the half-filled system. A negative value of $\Delta E$ ($\Delta E < 0$) indicates that it is energetically favorable for holes to form bound pairs. In contrast, $\Delta E \gtrsim 0$ signifies the absence of hole binding (note that $\Delta E$ may take slightly positive values in finite-size systems).
\begin{figure*}[!ht]
\hspace*{-0.5cm}
\vspace*{0cm}
\begin{overpic}[width=2.1\columnwidth]{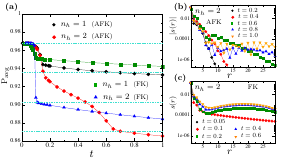}
\end{overpic}
\caption{{\bf Average plaquette operator and spin correlations as a function of hopping. } 
{\bf (a)} Average plaquette operator $P_{\mathrm{avg}}=\frac{1}{N_p}\sum_i \langle W_p(l) \rangle$ at the Kitaev points $\phi=\pm \pi/2$ as a function of the hopping amplitude $t$ for one ($n_h=1$) and two ($n_h=2$) doped holes. At half-filling $W_p$ is a good quantum number and the ground state has $P_\mathrm{avg}=1$. Deviations from this value thus capture deviations from the QSL state. 
Dotted lines indicate reference values of $P_{\mathrm{avg}}$ for $n$ broken plaquettes. The top line corresponds to $n=1$, with descending lines representing $n = 2, 3, 4$, respectively.
{\bf (b,c)} Spin–spin correlation function $S(r)$ for two doped holes at different values of $t$ in the {\bf (b)} AFK and {\bf (c)} FK regions.  
These numerical calculations were performed for a fixed system size $N=2\times63$.}
\label{fig3}
\end{figure*}

In Fig.~\ref{fig2}(a), we present the binding energy $\Delta E$ as a function of $\phi$ in the range $-\pi/2 \leq \phi \leq \pi/2$ for several system sizes, $N = 2 \times 23$, $2 \times 33$, $2 \times 53$, and $2 \times 63$, at a fixed hopping amplitude $t = 1$. The binding energy takes negative values only within the range $-0.29\pi \leq \phi \leq 0.48\pi$, indicating that hole pairing occurs only in the rung-singlet phase. This is consistent with our prior finding on three-leg ladders that hole pairing at $t=1$ occurs only within the N\'eel phase \cite{PhysRevB.110.224518}. In the rung-singlet region of the two-leg ladder, the values of the binding energy for different system sizes almost overlap, indicating that finite-size effects are negligible [Fig.~\ref{fig2}(a)]. In the rung-singlet phase, hole binding is energetically favored because it minimizes the number of broken singlet bonds; qualitatively two bound holes break only one rung singlet, whereas two unbound holes would break two. The positive binding energy observed in the stripy, AFK, and FK phases at $t = 1$ rules out the possibility of hole pairing for moderate hopping strengths, which is also consistent with our previous study on the three-leg honeycomb Kitaev–Heisenberg ladder model \cite{PhysRevB.110.224518}. The binding energy in Fig.~\ref{fig2}(a) exhibits an asymmetry around $\phi = 0$. 
This asymmetric behavior of the binding energy closely follows the behavior of the many-body excitation gap \cite{PhysRevB.99.224418} (see Supplementary Note 1 for details).

As discussed above, the binding energy is non-negative at the Kitaev points ($K = \pm 1$ and $J = 0$) for intermediate hopping strength $t = 1$. It is therefore interesting to examine how the kinetic energy of dopants influences hole pairing within the Kitaev spin-liquid phases for two leg-ladder $t$--$K$--$J$ model. In Fig.~\ref{fig2}(b,c), we show the binding energy for the AFK and FK spin-liquid phases as a function of the hopping parameter $t$. For the AFK phase, the binding energy becomes negative for $t \le t^c_\mathrm{AFK} \approx 0.65 K$ across all system sizes, indicating the formation of bound hole pairs in the slow-hole limit [Fig.~\ref{fig2}(b)]. In contrast, for the FK spin-liquid phase, hole binding occurs only at much lower hopping values, $t \le t^c_\mathrm{FK} \approx 0.1 |K|$ [Fig.~\ref{fig2}(c)]. The lower critical hopping $t^c$, in the FK phase compared to the AFK phase arises from the emergence of a kinetic ferromagnetic state in the FK regime even for small $t$ and  low hole concentrations; see  Fig.~\ref{fig3}(c). In the exactly solvable slow-hole limit of $t=0$ where the holes can be interpreted as static vacancies, the binding energies for both phases converge to the same value, $\Delta E = -0.25$, showing their independence from the sign of $K$. The hole binding observed in both AFK and FK phases as $t \to 0$ can be qualitatively understood as a consequence of minimizing the number of broken magnetic bonds due to hole pairing, following standard $t-J$ model arguments \cite{RevModPhys.66.763, PhysRevB.110.224518}.

\noindent\\ {\small \bf Effect of hole doping on spin-liquid phases\\}
Having found that the hopping amplitude $\frac{t}{|K|}$ controls the pairing tendencies at the Kitaev points, we next investigate how hole doping and the kinetic energy of dopants modify the AFK and FK spin-liquid backgrounds. 
We first focus on the one- and two-hole doping cases by calculating the average plaquette operator and spin–spin correlation functions. The Kitaev Hamiltonian ($K = \pm 1$, $J = t = 0$) possesses a conserved quantity called the plaquette operator $W_p$ \cite{Kitaev2006}, defined on each elementary hexagon $p$ of the ladder as
\begin{equation}
W_p = 2^6 S^x_1 S^y_2 S^z_3 S^x_4 S^y_5 S^z_6,
\label{eq:plaquette}
\end{equation}
where the subscripts $1, \dots, 6$ label the sites of the elementary plaquette $p$ as indicated in Fig.~\ref{fig1}(a). 
$W_p$ has eigenvalues $W_p = +1$ and $W_p = -1$ corresponding to the $Z_2$ gauge fluxes of $0$ and $\pi$, respectively. 
In the absence of hole doping, the ground state at either Kitaev point is flux free, with the expectation value of each plaquette operator given by $\langle W_p \rangle = 1$. 
Deviations from this value thus quantify the modification of the spin liquid ground state due to perturbations by doped holes or additional interactions \cite{Gordon2019, Kadow_2024, Jin2024}.

In Fig.~\ref{fig3}(a), we show the plaquette operator averaged over all hexagonal plaquettes of the two-leg ladder, defined as $P_{\mathrm{avg}} = \frac{1}{N_p} \sum_l \big\langle W_p(l)\big\rangle $, where $l$ labels the plaquette index and $N_p$ is the total number of plaquettes, as a function of the hopping amplitude $t$ for one- and two-hole doping. 
For $\frac{t}{|K|} \ll 1$, the holes remain confined to the boundary sites and act as quasi-static vacancies in both the FK and AFK spin-liquid phases. In the slow-hole limit ($t\ll |K|$), a single hole thus breaks only one plaquette near the boundary, yielding $P_{\mathrm{avg}} \approx 0.9677$. In this regime, we observe finite local magnetization, $\langle S_i^z \rangle$ or $\langle S_i^x \rangle$, localized near the edge sites where vacancies are created in both FK and AFK phases.  
Upon increasing $t$, a distinct kink appears in $P_{\mathrm{avg}}$ for the one-hole case, signaling a transition in which the hole relocates toward the center of the ladder and spreads over multiple sites. The characteristic crossover scale occurs at $t^{*}_\mathrm{FK}\!\approx 0.09|K|$ in the FK regime and $t^{*}_\mathrm{AFK}\!\approx 0.18K$ in the AFK regime.  In both regimes, $P_{\mathrm{avg}}$ remains finite and approaches $\approx 0.935$ at $\frac{t}{|K|}=1$, indicating that on average only two plaquettes are effectively destroyed by hole motion while the overall flux structure of the spin liquid largely survives. 
 
\noindent

Next, we examine the effect of two-hole doping in both AFK and FK phases as the hopping amplitude $t$ increases. In the AFK phase, the averaged plaquette operator $P_{\mathrm{avg}}$ decreases more rapidly with increasing $t$ [Fig.~\ref{fig3}(a)]. The $P_{\mathrm{avg}}$ curve shows two distinct slope changes at $t  \approx 0.19K$ and $t \approx 0.65K$, indicating two characteristic crossover scales. The second slope change is associated with the disappearance of hole pairing at $t^c_\mathrm{AFK} \approx 0.65K$, where the binding energy changes from negative to positive [see Fig.~\ref{fig3}(a) and Fig.~\ref{fig2}(b)]. In the FK case, the averaged plaquette operator $P_{\mathrm{avg}}$ for two-hole doping shows only one transition at $t \approx 0.1 |K|$, the same value as $t^c_\mathrm{FK}$ at which the binding energy changes from negative to non-negative in the FK regime [see Fig.~\ref{fig3}(a) and Fig.~\ref{fig2}(c)]. Notably, the average plaquette expectation value is larger in the FK case than in AFK [Fig.~\ref{fig3}(a)], showing that the underlying flux background is less disrupted by hole motion in FK (at least for the range of $t$ we studied).

To characterize the evolution of magnetic correlations, we calculate the  spin–spin correlation function along the upper leg of the ladder,
\begin{equation}
s(r) = \langle \mathbf{S}_{i_0} \cdot \mathbf{S}_j \rangle,
\end{equation}
where we choose the reference site $i_0 = 27$ near the center of the ladder, and define $r = \left| j - i_0\right|$ as the distance along the leg direction.
In the AFK phase, $|s(r)|$ decays exponentially for small $\frac{t}{K}$ [Fig.~\ref{fig3}(b)], consistent with short-range spin-liquid behavior. 
However, for $t > 0.65K$, $|s(r)|$ saturates to a finite value after an initial exponential decay, indicating the onset of magnetic correlations in the system. 
In contrast, in the FK phase magnetic correlations emerge already at small hopping, $t \sim 0.1|K|$ [Fig.~\ref{fig3}(c)], where $s(r)$ develops a quasi-long-range profile characteristic of a Nagaoka-type ferromagnetic state, in agreement with Ref.~\cite{Jin2024}. In this phase, the local spin moment acquires a small but finite value and exhibits a weak spatial modulation (with a two-peak structure along the ladder). This modulation is also reflected in the behavior of the spin-spin correlator $s(r)$, which displays the corresponding spatial variation with $r$ [see Fig.~\ref{fig3}(c)]. These transitions in spin-spin correlations are consistent with the binding energy results in the FK and AFK regimes, where $\Delta E$ changes from negative to positive at approximately the same crossover hopping amplitude $t^{*}=0.1|K|$ and $t^*=0.65K$, respectively. These results indicate close links between the binding energy, the emergence of spin-spin correlations and the deterioration of the spin liquid background.

\begin{figure}[!ht]
\hspace*{-0.5cm}
\vspace*{0cm}
\begin{overpic}[width=1.1\columnwidth]{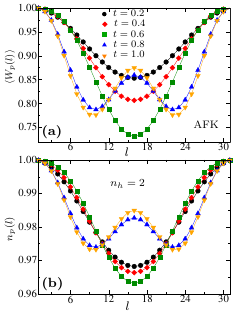}
\end{overpic}
\caption{{\bf Spatial correspondence between the local plaquette operator and charge-density profile in the two-hole doped system (AFK regime).} 
{\bf (a)} Spatial profiles of the local plaquette expectation value $\langle W_p(l) \rangle$ in the AFK limit, plotted for various hopping amplitudes $t$. {\bf (b)} Corresponding plaquette-averaged charge density $n_{p}(l)$ in the two-hole doped system, displayed for increasing hopping strength $t$. Notably, $\langle W_p(l) \rangle$ and $n_{p}(l)$ exhibit strong spatial congruence: a single minimum in the range of $t\leq 0.6K$ where pair formation occurs, and two minima for $t\geq 0.8K$ where doped holes repel instead of forming bound pairs. This correlation underscores that the dopants locally disturb the spin liquid background in the plaquettes around where the holes are delocalized. Data are shown for a fixed system size of $N = 2 \times 63$.}
\label{fig4}
\end{figure}

In Fig.~\ref{fig4}, we present the spatially resolved expectation value of the flux operator $\langle W_p(l) \rangle$ alongside the plaquette-averaged charge density, defined as $n_p(l)=\frac{1}{6}\sum_{j\in p_l}\langle n_j\rangle$, for the two-hole-doped case. Here $l$ labels the plaquette and the sum runs over the six sites $j$ belonging to plaquette $p_l$. The spatial profiles of the two quantities are strikingly similar. 
Interestingly, for smaller hopping ($t \lesssim 0.65K$), both $\langle W_p(l) \rangle$ [Fig.~\ref{fig4}(a,b)] and $n_p(l)$ exhibit a single broad minimum near the center, indicating that the two holes form an extended bound state that perturbs a contiguous region of plaquettes. Since the bound states can move, the width of the minimum does not directly correspond to the size of the bound state. 
The depth of the minimum increases with the hopping strength $\frac{t}{K}$, reflecting enhanced local plaquette distortion as the holes become more mobile. In contrast, for larger hopping ($t \gtrsim 0.65K$), $\langle W_p(l) \rangle$ and $n_p(l)$ develop two well-separated minima, showing that the holes move independently and disrupt distinct plaquette regions. The transition from a single-minimum (bound-pair) structure to a two-minima (unbound) structure in $\langle W_p(l) \rangle$ and $n_p(l)$  at $t\sim 0.65K$ is fully consistent with the pairing tendencies inferred from the binding-energy calculations.

\begin{figure*}[!ht]
\hspace*{-0.5cm}
\vspace*{0cm}
\begin{overpic}[width=2.1\columnwidth]{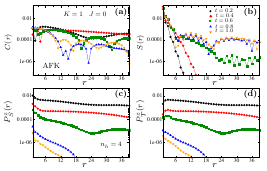}
\end{overpic}
\caption{{\bf Comparison of averaged correlation functions in the four-hole doped AFK regime with various hopping amplitudes.} 
{\bf (a)} Charge–charge correlations $C(r)$, {\bf (b)} spin–spin correlations $S(r)$,  
{\bf (c)} singlet pair–pair $P^{z}_{S}(r)$, and {\bf (d)} triplet pair–pair $P^{z}_{T}(r)$ correlations vs. distance $r$, plotted for representative values of the hopping amplitude $t$. The pair-pair correlations are strongly suppressed for $t\geq 0.8K$, consistent with the absence of pair formation. This is accompanied by enhanced spin-spin correlations, signaling a transition into a SDW phase.
Data correspond to a fixed system size of $N = 2 \times 63$.}
\label{fig5}
\end{figure*}

\noindent {\small \bf Comparison of charge, spin and  pairing order in the doped AFK spin liquid \\}
The above results show that, in the AFK phase, hole pairing occurs for $t \lesssim 0.65K$. 
However, the formation of bound hole pairs alone is not a sufficient condition for superconductivity, which also requires phase coherence among these pairs. 
To examine this, we compute both singlet and triplet pair–pair correlation functions and compare them with the corresponding charge–charge and spin–spin correlation functions. 
We find that the pair–pair correlations along the $x$- and $y$-bond directions either decay exponentially or remain very small in magnitude, indicating that hole pairing occurs predominantly along the rungs of the ladder (i.e., the $z$ bonds). Therefore, in the following we focus on the pair–pair correlation function with paring along the $z$-bond. 
The pair creation operator along the $i$th rung is given by
\begin{equation}
    \Delta^{\dagger}_{z}(i) = \frac{1}{\sqrt{2}} 
    \left[ 
    c^{\dagger}_{i,1,\uparrow} c^{\dagger}_{i,2,\downarrow} 
    \pm 
    c^{\dagger}_{i,1,\downarrow} c^{\dagger}_{i,2,\uparrow}
    \right],
\end{equation}
where the negative (positive) sign corresponds to singlet (triplet) pairing and $1,2$ are the ladder indices. The averaged pair-pair correlation for the $z$-bond is defined as
\begin{equation}
    P^{z}_{S/T}(r) = \frac{1}{N_r} \sum_{i,j} \delta_{|i-j|,r}
\Big\langle \Delta_{z}^{\dagger}(i)\, \Delta_{z}^{\phantom{\dagger}}(j) \Big\rangle ,
\end{equation}
where $N_r$ is the number of bond pairs separated by distance $r=j-i$ along the leg direction of the ladder. We also compute the averaged charge–charge correlation $C(r)$  along the upper leg, defined as
\begin{equation}
C(r) = \frac{1}{N_r} \sum_{i,j} \delta_{|i-j|,r}
\Big[
\langle n_i n_j \rangle - \langle n_i \rangle \langle n_j \rangle 
\Big],
\end{equation}
where $n_i$ denotes the local charge density at site $i$.  
The averaged spin–spin correlation is defined as
\begin{equation}
S(r) = \frac{1}{N_r} \sum_{i,j}  \delta_{|i-j|,r} \left| \langle \mathbf{S}_i \cdot \mathbf{S}_j \rangle \right|.
\end{equation}
To minimize boundary effects, we exclude $N = 2\times 11$ sites from each end of the ladder when computing the correlation functions.

\begin{figure*}[!ht]
\hspace*{-0.5cm}
\vspace*{0cm}
\begin{overpic}[width=2.1\columnwidth]{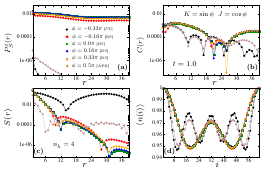}
\end{overpic}
\caption{{\bf Comparison of averaged correlation functions in different phases at fixed hopping and for four doped holes.}  
{\bf (a)} Singlet pair–pair correlations $P_{S}^{z}(r)$, {\bf (b)} charge–charge correlations $C(r)$, {\bf (c)} spin–spin correlations $S(r)$, and {\bf (d)} charge density $\langle n(i)\rangle$ along the leg direction, evaluated for $n_h = 4$ over the range $-0.33\pi \le \phi \le 0.5\pi$ at fixed hopping $t = 1$. The correlation functions and the charge density profile remain remarkably stable in the rung-singlet phase, which has robust pairing tendencies. In contrast, the AFK limit $(\phi = 0.5\pi)$ and the transition to the stripy phase ($-0.33\pi$) exhibit suppressed pairing and enhanced spin correlations. These regimes are further characterized by by a charge density profile with four distinct minima, signaling strong inter-hole repulsion that precludes the formation of local bound pairs. Data correspond to a fixed system size of $N = 2\times 63$.
}
\label{fig6}
\end{figure*}

Figure~\ref{fig5} compares the absolute values of the distance-averaged charge-charge correlation $C(r)$, spin-spin correlation $S(r)$, and singlet/triplet pair-pair correlations $P^{z}_{S/T}(r)$ (evaluated along the $z$ bonds) for $n_h=4$ holes as a function of the hopping amplitude $\frac{t}{K}$. For small hopping values ($t \lesssim 0.6K$), $S(r)$ decays exponentially, indicating the suppression of magnetic correlations, while $C(r)$ and $P^{z}_{S/T}(r)$ exhibit competing behavior. At $t = 0.2K$, the singlet and triplet pair–pair correlations slightly dominate over $C(r)$, suggesting the emergence of superconducting tendencies in the slow-hole limit. As $t$ increases to $t = 0.4K$ and $t = 0.6K$, the charge–charge correlations become dominant, signaling a crossover to charge–density–wave (CDW) ordering. 
For larger hopping ($t>0.6K$) [Fig.~\ref{fig5}(c,d)], the pair–pair correlations decay exponentially, while $C(r)$ and $S(r)$ remain comparable. In this regime, it is unclear whether the spin–spin or charge–charge correlations dominate at long distances. Moreover, for $n_h=4$ and $t_h \sim 1$ the spin and charge static structure factors exhibit only very broad features (see Fig.~7), we therefore refer to this regime as a disordered phase.
 Overall, these results demonstrate that hole doping and the kinetic energy of dopants drive a rich interplay between pairing, charge, and spin orders.
 
\noindent {\small \bf Comparison of charge, spin, and pairing correlations for finite $K$ and $J$\\}
Finally, we compare the charge-charge, spin-spin, and pair-pair correlations in the presence of both Kitaev ($K = \sin\phi$) and Heisenberg ($J = \cos\phi$) interactions. 
We vary the angle $\phi$ in the range $-0.33\pi \le \phi \le 0.5\pi$ and compute correlation functions for $n_h = 4$ holes at an intermediate hopping amplitude $t = 1$. 
The singlet pair-pair correlation $P^S_z$ dominates within the range $-0.27\pi \lesssim \phi \lesssim 0.44\pi$ [Fig.~\ref{fig6}(a)], consistent with the pairing tendencies inferred from the negative binding energy in the rung-singlet phase [Fig.~\ref{fig2}(a)]. 
In this rung-singlet regime, the spin-spin correlation $S(r)$ [Fig.~\ref{fig6}(c)] decays exponentially, whereas the charge-charge correlation $C(r)$ [Fig.~\ref{fig6}(b)] competes closely with the pair-pair correlation upon hole doping. For $\phi=-0.33\pi$ (stripy phase), the spin-spin correlations dominate, consistent with spin-density-wave order (SDW$_2$).    Figure~\ref{fig6}(d) shows the local charge density $\langle n_i \rangle$ as a function of site index $i$: in the rung-singlet phase, two pronounced minima appear, corresponding to two bound hole pairs (four holes in total). 
In contrast, for $\phi = -0.33\pi$ (stripy phase) and $\phi = 0.5\pi$, four distinct minima emerge, reflecting four unpaired and spatially separated holes in the system. 
These results are summarized in the doping-dependent phase diagram of Fig.~\ref{fig1}(f) along with results for many more values of $\phi$ and $n_h$ than shown here. Results for the $n_h$ dependence at representative $\phi$ values are shown in Supplementary Note 2.

\noindent\\ {\bf Discussion} \\
We have demonstrated that pairing tendencies at the Kitaev points (at least in the gapless, isotropic Kitaev model) are highly sensitive to the kinetic energy of the dopants, and that the pairing tendencies follow the same trends as the plaquette operator. The latter is a good quantum number only in the Kitaev limit, and its decay with increasing hopping directly reflects the breakdown of the quantum spin liquid state. Indeed, its spatial profile mimics the spatial charge distribution, namely the expectation value of the plaquette operator dips at the same locations where the dopants are preferentially located; see Fig.~\ref{fig4}. We interpret this as a signature of the dopants locally breaking the QSL background over a range set by their delocalization length. As the hopping strength is increased, this range increases, and eventually the QSL state deteriorates to a point where it fails to support pairing. This is reflected in the vanishing of the binding energy, in a doubling of the number of troughs in the spatially resolved plaquette operator and charge density, and remarkably also in kinks in the average value of the plaquette operator. We note that Ref. \cite{Jin2024} also found a kink in the average plaquette operator at the transition from QSL to FM state in the one-hole doped FK model. 

This strongly $t$-dependent behavior stands in stark contrast to the behavior at the antiferromagnetic Heisenberg point ($J=1$, $K=0$), where the binding energy is not sensitive to the hopping strength over the entire range we have considered here. This contrast in results agree with our previous findings for the three-leg ladder \cite{PhysRevB.110.224518} and match the more familiar behavior of $t-J$ models on lattices with square motifs \cite{RevModPhys.66.763, doi:10.1126/science.271.5249.618, RevModPhys.78.17}. The distinct sensitivity to the hopping strength at the Kitaev points is a signal that kinetic effects play a key role in destabilizing the QSL background and therefore obstructing pairing. Indeed, considering the results in Fig.~\ref{fig5} we see a marked suppression of pair-pair correlations as $t$ is raised, accompanied by the emergence of competing phases.

The distinction between slow ($t\ll 1$) and fast holes ($t\gg 1$) is fundamental to understanding doping effects in Kitaev systems. It explains the apparent conflict between early reports that a single hole in a Kitaev spin liquid propagates incoherently (for fast holes) \cite{PhysRevLett.111.037205,PhysRevB.90.024404} or coherently (for slow holes) \cite{PhysRevB.90.035145, PhysRevB.94.235105}. Recent numerical work has also revealed significant tendencies towards Nagaoka ferromagnetism \cite{Kadow_2024, Jin2024} that strengthen as $t$ increases.
Our plaquette operator results can thus be interpreted as follows. For slow holes, the quantum spin liquid background remains mostly undisturbed, and holes can propagate coherently, allowing them to form pairs. There is a crossover to incoherent hole transport above some critical hopping strength. Our results suggest this may be concomitant with the kinks in the plaquette operator, which indicate a sudden deleterious change to the spin liquid background, and which coincide with the disappearance of pair formation. 
Between prior work and our results herein and in our previous work \cite{PhysRevB.110.224518}, there is now robust numerical evidence for this distinction between slow and fast holes in cylinders of finite circumference. These observations are expected to generalize to the full two-dimensional honeycomb lattice, both because of the surprisingly close correspondence previously observed \cite{PhysRevB.99.195112, PhysRevB.99.224418, PhysRevX.11.011013} between Kitaev systems on quasi-one-dimensional geometries and in the 2D limit and because of the short-ranged nature of pairing processes and magnetic correlations in Kitaev systems.

We expect faster holes with $t\geq 1$ in realistic doped bulk quantum materials. Our results thus imply that materials close to the Heisenberg limit, such as YbCl$_3$ \cite{Sala2021} may become superconducting under hole-doping, but that materials in the Kitaev limit are unlikely to do so. However, as previously argued \cite{PhysRevB.110.224518}, the slow-hole regime may be accessible through heterostructures \cite{PhysRevLett.123.237201} or quantum simulation platforms \cite{Bohrdt2021, PRXQuantum.4.020329,  PhysRevLett.132.186501}. We also note that current Kitaev candidate materials such as RuCl$_3$ and Na$_2$IrO$_3$ are known to require additional interactions not included in the minimal Kitaev-Heisenberg model considered here. Their impacts on pairing tendencies and the phase diagram is left for future work.

Separately, the binding energy for $J>0$ and fixed finite $t$ is not symmetric with respect to $\phi=0$. This can be explained as being due to the stronger tendency towards Nagaoka ferromagnetism \cite{Kadow_2024, Jin2024} and weakened pairing for FM Kitaev interactions. 
The strongest binding energy is found near $\phi=0.1$, i.e. near the maximum of the many-body excitation gap \cite{PhysRevB.99.224418} (see also Supplementary Note 1). Since magnetization is not conserved away from $\phi=0,\pi$ the gap cannot be identified as a pure spin gap, but it is nevertheless likely to strongly screen the spin-spin correlations, which is linked with increased pairing tendencies in $t-J$ and Hubbard models.

We also find negative binding energy in a narrow range around the quantum phase transition between the zigzag and FM phases at $t=1$; see Supplementary Note 3. This is entirely unexpected and may be a frustration effect. However, at such intermediate hopping, the negative binding energy coincides with a spin-density wave order. Whether a superconducting phase can be stabilized in the slow-hole regime is left for future work. We note that a coupled cluster study \cite{PhysRevResearch.6.033168} recently proposed that the Kitaev-Heisenberg model at half-filling on the 2D honeycomb lattice may have a narrow phase between zigzag and FM, however we do not find any evidence for such a phase in our calculations. Indeed, the phase boundaries we find on long ladders are in excellent agreement with prior exact diagonalization results \cite{PhysRevB.99.195112}.

In conclusion, our results highlight the role of kinetic effects in doped Kitaev systems. It will be interesting in future work to consider extended Kitaev-Heisenberg models, which include additional interactions such as the off-diagonal $\Gamma$ interaction \cite{PhysRevX.11.011013, Soerensen2024}. More generally, it is relevant to ask whether other quantum spin liquids with bond-anisotropic interactions would also be highly sensitive to the hopping strength.

\noindent {\bf Methods} \\
We solve the Hamiltonian [Eq.~\eqref{eq:ham}] on a two-leg ladder using the density-matrix renormalization group (DMRG) method~\cite{PhysRevLett.69.2863,PhysRevB.48.10345}, implemented in the DMRG++ code~\cite{Alvarez2009} and the ITensor library~\cite{itensor,itensor-r0.3}. We consider ladder systems with $N=2\times L_x$ sites, up to $N=126$. 
The corresponding number of plaquettes, $N_p$, satisfies $N=2+4N_p$, and is an integer if $L_x$ is odd. 
Binding energies and all correlation functions are computed using DMRG++. We retain up to $m=7200$ DMRG states, yielding truncation errors $\epsilon \sim 10^{-8}$--$10^{-9}$. The plaquette operator and local charge density are computed using ITensor. 
For the ITensor calculations, we employed a maximum bond dimension of $m_{\max} = 5000$ and a singular value truncation threshold of $10^{-10}$.
\\

\noindent {\bf Data availability} \\
The data underlying this work will be made available in a Zenodo repository.\\

\noindent {\bf Code availability} \\
DMRG++ and ITensor are open-source software packages available at \url{https://github.com/g1257/dmrgpp} and \url{https://github.com/ITensor}, respectively.\\

\noindent {\bf References} \\
%

\begin{acknowledgments}
The work of B.P., S.O., and E.D. was supported by the U.S. Department of Energy (DOE), Office of Science, Basic Energy Sciences (BES), Materials Sciences and Engineering Division. The work of B.X., G.A., and G.B.H. was supported by the U.S. Department of Energy, Office of Science, National Quantum Information Science Research Centers, Quantum Science Center. Parts of the computations for this work were performed on the high performance computing infrastructure operated by Research Support Solutions in the Division of IT at the University of Missouri, Columbia MO DOI: \url{https://doi.org/10.32469/10355/97710}.

This manuscript has been co-authored by UT-Battelle, LLC, under contract DE-AC05-00OR22725 with the U.S. Department of Energy (DOE).
The U.S. government retains and the publisher, by accepting the article for publication, acknowledges that the U.S. government retains a nonexclusive, paid-up, irrevocable, worldwide license to publish or reproduce the published form of this manuscript, or allow others to do so, for U.S. government purposes. DOE will provide public access to these results of federally sponsored research in accordance with the DOE Public
Access Plan \url{https://www.energy.gov/doe-public-access-plan}.\\
\end{acknowledgments}

\newpage
\clearpage
\title{Supplemental material for ``Luther-Emery liquid and dominant singlet superconductivity in the hole-doped Haldane spin-1 chain''}
\maketitle
\onecolumngrid

\section*{Supplementary Note 1: Phase boundaries and excitation gaps at half-filling}
\begin{figure*}[!ht]
	\hspace*{-0.5cm}
	\vspace*{0cm}
	\begin{overpic}[width=1.0\columnwidth]{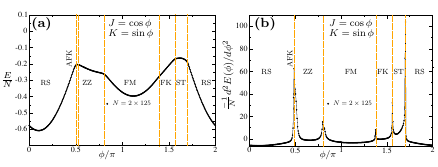}
	\end{overpic}
	\caption{\textbf{Phase boundary at half filling for $0 \le \phi \le 2\pi$.}
		(a) Ground state energy per site, $E$, as a function of $\phi$ in the second quadrant ($\pi/2 \le \phi \le \pi$) at fixed $t=1.0$.
		(b) Second derivative of the energy with respect to $\phi$, $-(1/N)\,d^{2}E(\phi)/d\phi^{2}$, as a function of $\phi$.
		All results are obtained for system size $N=250$.}
	\label{SFig1}
\end{figure*}
Figure \ref{SFig1} shows the ground state energy (a) and its second derivative (b) as a function of $\phi$ for the half-filled Kitaev-Heisenberg model. The results are in close agreement with the ED results on 24-site ladders with PBC previously reported in the Supplemental Material of Ref.~\cite{PhysRevB.99.195112}.

\begin{figure}
	\includegraphics[width=0.55\textwidth]{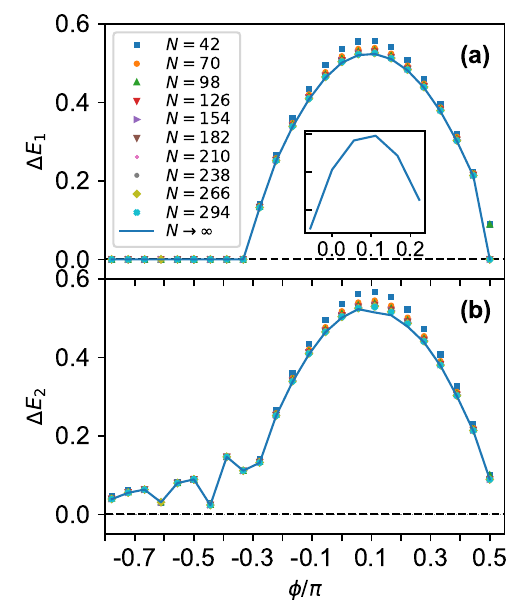}
	\caption{\textbf{Many-body excitation gaps.} \textbf{(a)} Gap between the first excited state and the ground state. The symbols shows calculated gaps from DMRG for for various system sizes $N=2\times L_x$. The solid line is a linear extrapolation to the thermodynamic limit. The inset shows a zoomed-in view of the peak. \textbf(b) Gap between the second excited state and the ground state.}
	\label{SFig2}
\end{figure}

We have calculated the many-body excitation gap in the half-filled Kitaev-Heisenberg model using DMRG. The gap between the $m$th excited state and the ground state is defined for fixed system size $N$ as
\begin{equation}
	\Delta E_m (N)  = E_m(N) -E_0(N),
\end{equation}
where $E_0$ is ground state energy and $E_m$ is the $m$th excited state energy. The results for $m=1,2$ are shown in Fig.~\ref{SFig2}, along with the extrapolation $\lim_{N\rightarrow\infty} \Delta E_m$. Our extrapolated results for $\Delta E_1$ are in excellent agreement with those reported in Ref.~\cite{PhysRevB.99.224418} for the range $\frac{\phi}{\pi}\in [-0.3, 0.5]$, and confirm that the gap is (i) asymmetric around $\phi=0$ and (ii) maximized away from $\phi=0$. 
Unlike the rung singlet phase, the ZZ, ST, AFK, and FK phases all have twofold degenerate ground states \cite{PhysRevB.99.195112} so the physical excitation gap in these phases is given by $\Delta E_2$.

\section*{Supplementary Note 2: Characterization of phases away from half-filling}

\subsection*{Comparison of correlation functions in the rung-singlet phase with increasing hole doping}

\begin{figure*}[!ht]
	\hspace*{-0.5cm}
	\vspace*{0cm}
	\begin{overpic}[width=1.0\columnwidth]{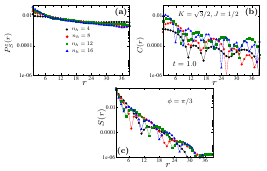}
	\end{overpic}
	\caption{\textbf{Comparison of averaged correlation functions at $\phi=0.33\pi$ for increasing hole number.}
		\textbf{(a)} singlet pair-pair correlations $P^{zz}_{S}(r)$,
		\textbf{(b)} charge-charge correlations $C(r)$, and
		\textbf{(c)} spin-spin correlations $S(r)$,
		shown as a function of distance $r$ for $n_h=4,8,12,$ and $16$ holes.
		The pair-pair correlations dominate for all studied $n_h$.
		The charge--charge correlations also exhibit a power-law decay and compete with the pair--pair correlations,
		whereas the spin-spin correlations decay exponentially for all $n_h$.
		All results are obtained for a fixed system size $N=2\times63$ and $t=1.0$.}
	\label{SFig3}
\end{figure*}

In Fig.~\ref{SFig3}, we compare the charge-charge, spin-spin, and pair-pair correlations in the rung-singlet phase at fixed $\phi=\pi/3$ (i.e., $K=\sqrt{3}/2$ and $J=1/2$) and $t=1.0$ for different hole numbers, $n_h=4,8,12,$ and $16$. As shown in Fig.~\ref{SFig3}(a), the singlet pair-pair correlation $P^{zz}_{S}(r)$ decays slowly with distance and is consistent with an algebraic (power-law-like) behavior for all doping levels considered. The charge-charge correlation $C(r)$ [Fig.~\ref{SFig3}(b)] also decreases with distance, but its decay is noticeably faster than that of $P^{zz}_{S}(r)$. In contrast, the spin-spin correlation $S(r)$ is strongly suppressed at long distances and remains short-ranged for all $n_h$ [Fig.~\ref{SFig3}(c)]. With increasing hole doping, we observe a modest enhancement of $C(r)$, while $P^{zz}_{S}(r)$ remains the dominant long-distance correlation, indicating that charge and superconducting correlations compete but that superconducting correlations continue to dominate in this parameter regime over the doping levels studied here.

\subsection*{Comparison of correlation functions in the stripy phase with increasing hole doping}

\begin{figure*}[!ht]
	\hspace*{-0.5cm}
	\vspace*{0cm}
	\begin{overpic}[width=1.0\columnwidth]{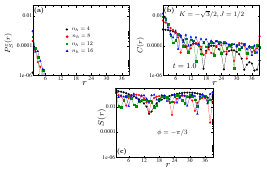}
	\end{overpic}
	\caption{\textbf{Comparison of averaged correlation functions at $\phi=-0.33\pi$ for increasing hole number.}
		\textbf{(a)} singlet pair-pair correlations $P^{zz}_{S}(r)$,
		\textbf{(b)} charge-charge correlations $C(r)$, and
		\textbf{(c)} spin-spin correlations $S(r)$,
		shown as a function of distance $r$ for $n_h=4,8,12,$ and $16$ holes.
		The pair-pair correlations decay exponentially for all studied $n_h$.
		The charge-charge correlations exhibit a power-law decay and compete with the spin-spin correlations,
		whereas the spin-spin correlations dominates for all $n_h$.
		All results are obtained for a fixed system size $N=2\times63$ and $t=1.0$.}
	\label{SFig4}
\end{figure*}

In Fig.~\ref{SFig4}, we compare the charge-charge, spin-spin, and pair-pair correlations in the stripy phase at fixed $\phi=-\pi/3$ (i.e., $K=-\sqrt{3}/2$ and $J=1/2$) and $t=1.0$ for different hole numbers, $n_h=4,8,12,$ and $16$. Consistent with the fact that the binding energy remains positive in this regime, the singlet pair-pair correlation $P^{zz}_{S}(r)$ is short-ranged and decays rapidly with distance [Fig.~\ref{SFig4}(a)]. By contrast, the charge-charge correlation $C(r)$ exhibits a slow (power-law-like) decay with $r$ for all doping levels shown [Fig.~\ref{SFig4}(b)]. The spin-spin correlation $S(r)$ decays most slowly among the three correlations [Fig.~\ref{SFig4}(c)], indicating that spin ordering is robust against hole doping in the stripy phase and remains quasi-long-ranged over the range of $n_h$ studied here.

\subsection*{Charge-charge and spin-spin correlations in the antiferromagnetic Kitaev spin-liquid phase with increasing hole doping}

\begin{figure*}[!ht]
	\hspace*{-0.5cm}
	\vspace*{0cm}
	\begin{overpic}[width=1.0\columnwidth]{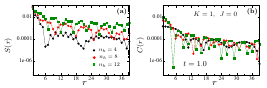}
	\end{overpic}
	\caption{\textbf{Comparison of averaged correlation functions at $\phi=\pi/2$ for increasing hole number.}
		\textbf{(a)} Spin-spin correlations $S(r)$ and \textbf{(b)} charge-charge correlations $C(r)$
		are shown as a function of distance $r$ for $n_h=4,8,12$ holes.
		The spin-spin correlations are enhanced with increasing hole doping. The charge-charge correlations also decay as a power law and compete with the spin-spin correlations. However, at higher doping levels the spin-spin correlations dominate over the charge-charge correlations. 
		All results are obtained for a fixed system size $N=2\times63$ and $t=1.0$.}
	\label{SFig5}
\end{figure*}

In Fig.~\ref{SFig5}, we examine the evolution of the spin-spin correlation $S(r)$ and the charge-charge correlation $C(r)$ in the antiferromagnetic Kitaev spin-liquid regime at a fixed intermediate hopping $t=1.0$ as the hole number is increased. At low doping, $S(r)$ and $C(r)$ decay with comparable rates, indicating a close competition between spin and charge correlations. Upon further hole doping, $S(r)$ is enhanced at long distances and decays more slowly, whereas $C(r)$ shows a similar overall decay behavior across the doping levels considered. In addition, the pair-pair correlations remain short-ranged (not shown), indicating the absence of dominant superconducting correlations in this parameter regime. Overall, these results suggest that sufficient hole doping strengthens magnetic ordering in the spin-liquid phase, leading to spin correlations that dominate the long-distance behavior.

\subsection*{Charge-density-wave order near the ST-RS transition.}
\begin{figure*}[!ht]
	\hspace*{-0.5cm}
	\vspace*{0cm}
	\begin{overpic}[width=1.0\columnwidth]{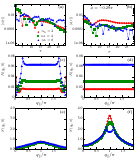}
	\end{overpic}
	\caption{\textbf{Charge-density-wave correlations near the ST-RS transition.}
		(a) Spin-spin correlations $|s(r)|$ and (b) charge-charge correlations $|c(r)|$, measured from the central site to sites on the right leg, plotted versus distance $r$ for $n_h=2,4,8$ at $\phi=-0.29\pi$.
		(c) Static charge structure factor $N(q_\parallel,0)$ and (d) $N(q_\parallel,\pi)$ plotted versus $q_\parallel$. 
		(e) Static spin structure factor $S(q_\parallel,0)$ and (f) $S(q_\parallel,0)$ plotted versus $q_\parallel$.
		All results are obtained for a fixed system size $N=2\times 63$ and $t=1.0$.}
	\label{SFig7}
\end{figure*}

Near the transition between the stripy (ST) and rung-singlet (RS) phases, around $\phi \simeq -0.29\pi$, we observe signatures of intertwined charge, spin, and pairing correlations.
In Fig.~\ref{SFig7}(a,b), we compare spin-spin and charge-charge correlation functions along the upper leg of the ladder, defined as
\begin{equation}
	s(r)=\langle \mathbf{S}_{i_0}\cdot \mathbf{S}_{j}\rangle,
\end{equation}
\begin{equation}
	c(r)=\langle n_{i_0} n_{j}\rangle-\langle n_{i_0}\rangle\langle n_{j}\rangle,
\end{equation}
where we choose a reference site $i_0=27$ near the center of the ladder and define the distance along the leg direction as $r=|j-i_0|$.
For $n_h<8$, the charge correlations $|c(r)|$ decay more slowly than the spin correlations $|s(r)|$, whereas for $n_h\ge 8$ the spin correlations dominate.
In Fig.~\ref{SFig7}(c,d), we compute the static charge structure factor
\begin{equation}
	N(q_\parallel,q_\perp)=\frac{1}{N}\sum_{i,j} e^{-i\mathbf{q}\cdot(\mathbf{r}_i-\mathbf{r}_j)}
	\left[\langle n_i n_j\rangle-\langle n_i\rangle\langle n_j\rangle\right],
	\qquad \mathbf{q}=(q_\parallel,q_\perp),
\end{equation}
where $N=2L$ is the total number of sites, $n_i$ is the density operator on site $i$, and $\mathbf{r}_i$ denotes the position of site $i$.
The structure factor $N(q_\parallel,0)$ exhibits two clear peaks, indicating incommensurate charge-density-wave behavior close to the ST-RS phase boundary.

We also compute the static spin structure factor
\begin{equation}
	S(q_\parallel,q_\perp)=\frac{1}{N}\sum_{i,j} e^{-i\mathbf{q}\cdot(\mathbf{r}_i-\mathbf{r}_j)}
	\left\langle \mathbf{S}_i \cdot \mathbf{S}_j \right\rangle,
	\qquad \mathbf{q}=(q_\parallel,q_\perp),
\end{equation}
where $\mathbf{S}_i$ is the spin operator on site $i$.
As shown in Fig.~\ref{SFig7}(e,f), $S(q_\parallel,0)$ displays a broad bump around $q_\parallel=\pi$, whereas $S(q_\parallel,\pi)$ shows a sharper peak at $q_\parallel=\pi$ for low hole doping.
For $n_h\ge 8$, $S(q_\parallel,\pi)$ develops incommensurate peaks, similar to the behavior of the static spin structure factor in the RS phase [see Fig.~\ref{SFig8}(e,f)].
Furthermore, the pair-pair correlations become closer to a power-law decay near the ST-RS transition, in contrast to the more rapid (exponential) decay in the stripy phase.
Taken together, these results suggest intertwined charge, spin, and pairing tendencies near the ST--RS phase boundary. Because the charge correlations dominate the long-distance behavior for $n_h<8$, we refer to this regime as exhibiting charge-density-wave (CDW) order in the main text.

\subsection*{Static spin structure factors}
\begin{figure*}[!ht]
	\centering
	\begin{overpic}[width=0.9\textwidth]{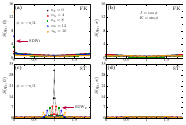}
	\end{overpic}
	\vspace{-3.0mm}
	\hspace*{-1.5mm}
	\begin{overpic}[width=0.9\textwidth]{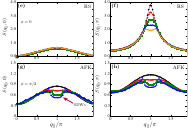}
	\end{overpic}
	\caption{\textbf{Static spin structure factor.}
		Static spin structure factors $S(q_x,0)$ and $S(q_x,\pi)$ plotted versus $q_x/\pi$ for different hole numbers $n_h$.
		Panels (a,c,e,g) show $S(q_x,0)$ for the FK phase at $\phi=-\pi/2$, the ST phase at $\phi=-\pi/3$, the RS phase at $\phi=0$, and the AFK phase at $\phi=\pi/2$, respectively.
		Panels (b,d,f,h) show $S(q_x,\pi)$ for the same phases and $\phi$ values, respectively. All results are obtained for a fixed system size $N=2\times 63$ and $t=1.0$. The purple arrows indicate the spin-density-wave phases SDW$_1$,  SDW$_2$, and  SDW$_3$. }
	\label{SFig8}
\end{figure*}
	To further clarify the nature of the doped phases, we calculate the static spin structure factor $S(q_\parallel,q_\perp)$ for representative values of $\phi=-\pi/2$ (FK), $-\pi/3$ (ST), $0$ (RS), and $\pi/2$ (AFK) (see Fig. \ref{SFig8}).
	In the FK limit, a strong $S(q_\parallel,0)$ peak develops at $q_\parallel=0$ with finite doping, signaling Nagaoka ferromagnetism [Fig. \ref{SFig8} (a,b)]. We denote this commensurate spin-density wave order SDW$_1$ in the main text. At half-filling in the stripy phase, the AFM order along the legs results in a strong $S(q_\parallel,0)$ peak at $q_\parallel=\pi$ [Fig. \ref{SFig8} (c,d)]. At finite doping the single peak at $q_\parallel=\pi$ splits into two incommensurate peaks appearing symmetrically around $q_\parallel=\pi$, as is typical for $t-J$ and Hubbard two-leg ladders. As previously shown the spin-spin correlations dominate the long-range behavior at this value of $\phi$ [Fig. \ref{SFig4}]. We denote this incommensurate spin-density-wave order SDW$_2$ in the main text. In the rung-singlet phase, $S(q_\parallel,0)$ exhibits a broad bump around $q_\parallel=\pi$, whereas $S(q_\parallel,\pi)$ shows a peak at $q_\parallel=\pi$ at half filling ($n_h=0$) [Fig. \ref{SFig8} (e,f)]. This behavior is consistent with short-range antiferromagnetic correlations along the legs and antiferromagnetic ordering across the rungs. Upon hole doping, the peak at $q_\parallel=\pi$ splits into two incommensurate peaks.
	In the AFK limit both $S(q_\parallel,0)$ and $S(q_\parallel,\pi)$ show very broad features at half-filling and for $n_h=4$ [Fig. \ref{SFig8} (g,h)]. Since it is also unclear from Figure \ref{SFig5} whether spin-spin or charge-charge correlations dominate at large distance (both are much larger than the pair-pair correlations, as shown in Fig. 5 of the main text) we call this a disordered phase. At $n_h\geq 12$ clear incommensurate peaks appear in $S(q_\parallel,0)$, similar to the behavior in the stripy phase. We denote this incommensurate spin-density wave order $SDW_3$ in the main text.

\clearpage
\section*{Supplementary Note 3: Binding energy for \texorpdfstring{$\pi/2\leq \phi \leq \pi$}{pi/2<phi<pi}}
\begin{figure*}[!ht]
	\hspace*{-0.5cm}
	\vspace*{0cm}
	\begin{overpic}[width=0.8\columnwidth]{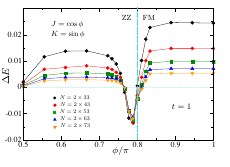}
	\end{overpic}
	\caption{\textbf{Binding energy for $\pi/2 \le \phi \le \pi$.}
		Binding energy $\Delta E$ as a function of $\phi$ in the second quadrant ($\pi/2 \le \phi \le \pi$) at fixed $t=1.0$.
		$\Delta E$ becomes negative in a narrow window, $0.75\pi \le \phi \le 0.81\pi$, close to the ZZ--FM phase boundary.
		The ZZ--FM transition occurs at $\phi \approx 0.8\pi$ and is indicated by the vertical cyan dotted line. Results are obtained for system sizes $N=2\times 33, 2\times 44, 2\times 53, 2\times 64,$ and $2\times 73$.}
	\label{SFig6}
\end{figure*}

In Fig.~\ref{SFig6}, we show the binding energy $\Delta E$ as a function of 
$\phi$ in the range $\pi/2 \le \phi \le \pi$ for several system sizes, $N=2\times33$, $2\times43$, $2\times53$, $2\times63$, and $2\times73$, at a fixed hopping amplitude $t=1.0$. At half filling, the ground state in this quadrant evolves from the zigzag (ZZ) phase into the ferromagnetic (FM) phase, with the ZZ-FM transition occurring near $\phi \simeq 0.8\pi$. We find that $\Delta E$ becomes negative in a narrow window, $0.75\pi \lesssim \phi \lesssim 0.81\pi$, close to the ZZ-FM phase boundary, indicating a weak tendency toward pairing in this region. However, the magnitude of $\Delta E$ is small compared to that in the rung-singlet regime, and the corresponding pair-pair correlations remain short-ranged, whereas the spin-spin and charge-charge correlations are comparatively stronger. Overall, for $\pi/2 \le \phi \le \pi$ and for the doping levels considered in this work ($n_h \le 16$), the long-distance behavior is dominated by spin--spin correlations, indicating robust magnetic correlations upon hole doping in this parameter regime. The negative binding energy seen in this range may thus be a frustration effect associated with the quantum phase transition.

\end{document}